\RequirePackage[2020-02-02]{latexrelease}
\documentclass[twocolumn,preprintnumbers]{revtex4}%
\usepackage{amsopn}
\usepackage{amssymb}
\usepackage{amsmath}
\usepackage{units}
\usepackage{graphicx}
\usepackage{dcolumn}
\usepackage{bm}
\usepackage{xcolor}
\usepackage{ifthen}
\usepackage[normalem]{ulem}
\usepackage{amsfonts}%
\RequirePackage[2020-02-02]{latexrelease}
\UseRawInputEncoding
\providecommand{\U}[1]{\protect\rule{.1in}{.1in}}
\providecommand{\U}[1]{\protect\rule{.1in}{.1in}}
\def\showal{1}
\newcommand{\al}[1]{\ifthenelse{\showal=1}{\textcolor{orange}{[[#1]]}}{}}

\newcommand{\eb}[1]{\ifthenelse{\showal=1}{\textcolor{cyan}{[[#1]]}}{}}

\begin{document}
\title{Intermode coupling in a fiber loop laser at low temperatures}
\author{Eyal Buks}
\email{eyal@ee.technion.ac.il}
\affiliation{Andrew and Erna Viterbi Department of Electrical Engineering, Technion, Haifa
32000, Israel}
\date{\today }

\begin{abstract}
We experimentally study an unequally-spaced optical comb (USOC),
which is generated by a unidirectional fiber loop laser operated at low
temperatures. The underlying mechanism responsible for USOC formation is
explored using both close and open loop measurements. The role played by
dispersion is investigated using radio frequency spectrum measurements. By integrating a saturable absorber into the loop, a lasing
state is revealed, in which mode locking coexists with the USOC.

\end{abstract}
\maketitle

\begin{figure}[ptb]
\begin{center}
\includegraphics[width=3.1in,keepaspectratio]{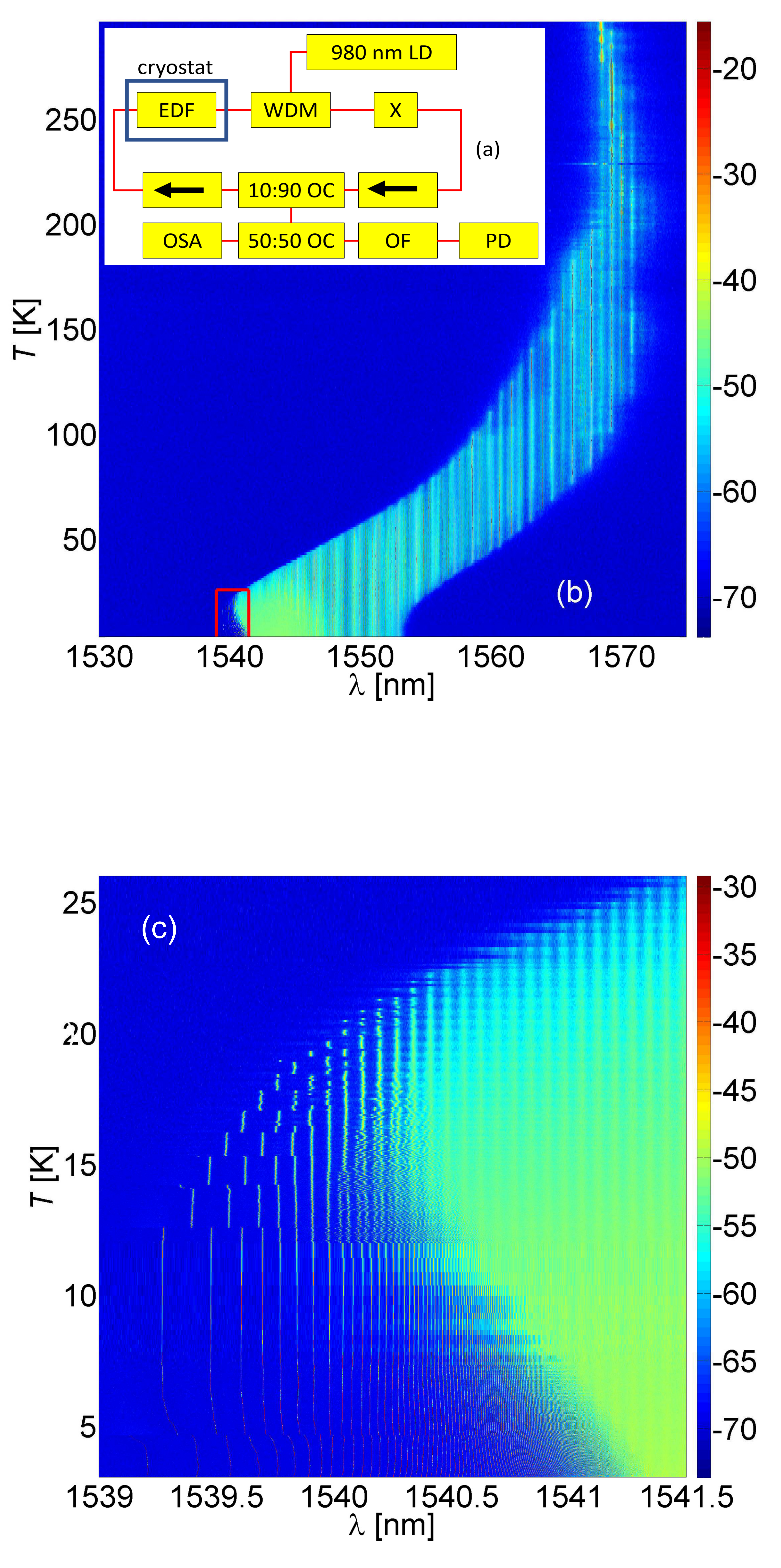}
\end{center}
\caption{{}Temperature dependency. (a) The experimental setup. The X box
represents a tunable attenuator, which is used for the measurements presented
in Fig. \ref{FigAtt}, a wavelength tunable OF, which is used for the
measurements presented in Fig. \ref{FigILOF}, and a SA, which is used for the
measurements presented in Fig. \ref{FigSA_G}. Spectrum measurements presented
in (b) and (c) are performed during cooling down. OSA data (here and in all
other figures) are presented in dBm units. The EDF length is $l_{\mathrm{EDF}%
}=5 \operatorname{m}$, and the diode current is (b) $I_{\mathrm{D}%
}=200\operatorname{mA}$ and (c) $I_{\mathrm{D}}=120\operatorname{mA}$.}%
\label{FigSetupT}%
\end{figure}

\textbf{Introduction} - Fiber loop lasers are widely employed for a variety of
applications. Commonly the loop is made of an undoped single mode fiber (SMF)
section, and a doped section, which is used for optical amplification. An
unequally-spaced optical comb (USOC) has been recently found in the optical
spectrum of a fiber loop laser having an amplifying Erbium doped fiber (EDF),
which is cooled down to cryogenic temperatures \cite{Buks_128591}. The underlying mechanism responsible for USOC formation has remained mainly
unknown. Here we report on measurements, performed with both
open and close loop configurations, which provide some insight into the process of USOC formation, and which allow the extraction of parameters of the fiber loop under study.

A SMF (both doped and undoped) can be characterized by a loss coefficient
$\alpha$, a group velocity $v_{\mathrm{g}}$, a group velocity dispersion
(GVD), and by a complex nonlinear coefficient $\gamma=\gamma^{\prime}%
+i\gamma^{\prime\prime}$, which is proportional to the SMF third-order
susceptibility \cite{Agrawal_Nonlinear_fiber_optics}. The real part
$\gamma^{\prime}$ of $\gamma$ is the nonlinear dispersion coefficient
\cite{Montagna_871}, and the imaginary part $\gamma^{\prime\prime}$ of
$\gamma$ is the nonlinear absorption coefficient
\cite{Jachpure_024007,Maeda_1187}. For saturable absorption (SA)
$\gamma^{\prime\prime}<0$, whereas $\gamma^{\prime\prime}>0$ for reverse
saturable absorption (RSA). In a fiber loop laser, SA commonly promotes mode
locking (ML) and the formation of optical pulses
\cite{Kapitula_740,Dennis_1469,Huang_034009}, whereas processes giving rise to
RSA (e.g. two-photon absorption) suppress ML \cite{Wang_106805}. Changeover
from SA to RSA has been observed in \cite{Wei_7704}.

Fiber parameters are extracted in the current study using a variety of
measurement techniques. The effect of EDF emission and absorption spectra is
studied by varying the temperature and fiber loop loss. The effect of nonlinear
dispersion is explored by measuring both the loop frequency as a function of
optical wavelength and the lasing linewidth. Both EDF gain and nonlinear
response are measured using an open loop configuration. We show that some of
the experimental results can be accounted for using a simple theoretical
model, which assumes that dispersion can be disregarded. However, on the other
hand, this assumption is invalidated by some of our experimental results,
which indicate that dispersion plays an important role.

\textbf{Experimental setup} - The experimental setup is described by the
sketch shown in Fig. \ref{FigSetupT}(a). A cryogen free cryostat is used to
cool down an EDF having length denoted by $l_{\mathrm{EDF}}$, absorption of
$30$ dB $%
\operatorname{m}%
^{-1}$ at $1530%
\operatorname{nm}%
$, and mode field diameter of $6.5%
\operatorname{\mu m}%
$ at $1550%
\operatorname{nm}%
$. The EDF, which is thermally coupled to a calibrated silicon diode
thermometer, is pumped using a $980%
\operatorname{nm}%
$ laser diode (LD) biased with current denoted by $I_{\mathrm{D}}$. Results
are presented below for two fiber loops, one has $l_{\mathrm{EDF}}=10%
\operatorname{m}%
$, and the other one has $l_{\mathrm{EDF}}=5%
\operatorname{m}%
$.

The cold EDF is integrated with a room temperature fiber loop using a
wavelength-division multiplexing (WDM) device. A 10:90 output coupler (OC),
and two isolators [labeled by arrows in the sketch shown in Fig.
\ref{FigSetupT}(a)], are integrated in the fiber loop. The output port of the
10:90 OC is splitted using a $50:50$ OC to allow probing the optical signal
using both an optical spectrum analyzer (OSA) and a photodetector (PD), which
is connected to either a radio frequency spectrum analyzer (RFSA), or to an
oscilloscope. An optional tunable optical filter (OF) having a central
wavelength denoted by $\lambda_{\mathrm{F}}$, and a linewidth of
$\delta_{\mathrm{F}}=1.2%
\operatorname{nm}%
$ (full width at half maximum) is connected between the $50:50$ OC and the PD.

\textbf{Temperature dependency} - Key properties of EDF can be
controlled by varying the temperature
\cite{Kobyakov_1,Le_3611,Thevenaz_22,Aubry_2100002,Pizzaia_2352,Liu_102988,Lopez_085401,Antuzevics_1149,Chu_966}%
. EDF operating at low temperatures can be used for some applications,
including multimode lasing \cite{perez2013multi,haken1985laser} and quantum
information storage
\cite{Saglamyurek_83,Staudt_720,Wei_2209_00802,Ortu_035024,Liu_2201_03692,Veissier_195138,Saglamyurek_241111,Shafiei_F2A,Buks_L051001}%
.

The measured optical spectrum of our device as a function of the temperature
$T$ with diode current of $I_{\mathrm{D}}=200%
\operatorname{mA}%
$ is shown in Fig. \ref{FigSetupT}(b). The wavelength at which lasing peaks is
denoted by $\lambda_{\mathrm{g}}$. The spectrum shown in Fig. \ref{FigSetupT}
indicates that $\lambda_{\mathrm{g}}$ decreases as the temperature is lowered
\cite{Kagi_261}. As is discussed below, this process is attributed to the
temperature dependency of the EDF emission and absorption spectra.

Following Ref. \cite{Franco_1090}, consider an infinitesimal EDF section of
length $\mathrm{d}z$. The signal gain along this section is expresses as
$1+\kappa\mathrm{d}z$, where $\kappa=l_{\mathrm{E}}^{-1}\left(
1-p_{\mathrm{g}}\right)  -l_{\mathrm{A}}^{-1}p_{\mathrm{g}}$, $l_{\mathrm{E}%
}^{-1}$ ($l_{\mathrm{A}}^{-1}$) is the emission (absorption) inverse length
(which may depend on both wavelength $\lambda$ and temperature $T$),
$p_{\mathrm{g}}$ is the ground state population fraction, and $1-p_{\mathrm{g}%
}$ is the excited state population fraction (note that $0\leq p_{\mathrm{g}%
}\leq1$, and that due to the isotropic nature of spontaneous emission, which
yields weak coupling to the fiber propagating mode, its contribution to the
signal is disregarded). The total EDF gain $g_{\mathrm{F}}$ is given by
$g_{\mathrm{F}}=\exp\int_{0}^{l_{\mathrm{EDF}}}\kappa\mathrm{d}z\;$. The
evolution of pump intensity $I_{\mathrm{P}}$ along the EDF is governed by
$\mathrm{d}I_{\mathrm{P}}/\mathrm{d}z=-l_{\mathrm{P}}^{-1}p_{\mathrm{g}%
}I_{\mathrm{P}}$, where $l_{\mathrm{P}}^{-1}$ is the inverse pump decay
length, thus $g_{\mathrm{F}}$ can be expressed as $g_{\mathrm{F}}=\exp\left(
l_{\mathrm{EDF}}/l_{\mathrm{E}}+l_{\mathrm{f}}^{-1}\zeta_{\mathrm{P}}\right)
$, where $l_{\mathrm{f}}^{-1}=l_{\mathrm{E}}^{-1}+l_{\mathrm{A}}^{-1}$, and
where $\zeta_{\mathrm{P}}=l_{\mathrm{P}}\mathrm{\log}\left(  I_{\mathrm{P}%
}\left(  l_{\mathrm{EDF}}\right)  /I_{\mathrm{P}}\left(  0\right)  \right)  $.
The total loop gain $g$ is expressed as $g=g_{\mathrm{F}}/\alpha_{\mathrm{FL}%
}$, where $\alpha_{\mathrm{FL}}\geq1$ is the fiber loop loss coefficient. The
lasing wavelength $\lambda_{\mathrm{g}}$ is determined from the conditions
that $\log g=0$ (i.e. $g=1$) and $\mathrm{d}\log g/\mathrm{d}\lambda=0$. By
eliminating the term $\zeta_{\mathrm{P}}$ from these two conditions one finds
that $\mathrm{d}\varrho/\mathrm{d}\lambda=0$ at the lasing wavelength
$\lambda=\lambda_{\mathrm{g}}$, where $\varrho$ is given by
\begin{equation}
\varrho=\left(  \eta_{\mathrm{EDF}}-1\right)  l_{\mathrm{f}}\log
\alpha_{\mathrm{FL}}\;, \label{varrho}%
\end{equation}
and where $\eta_{\mathrm{EDF}}=l_{\mathrm{EDF}}/\left(  l_{\mathrm{E}}%
\log\alpha_{\mathrm{FL}}\right)  $ is the effective and normalized EDF length.

\begin{figure}[ptb]
\begin{center}
\includegraphics[width=3.2in,keepaspectratio]{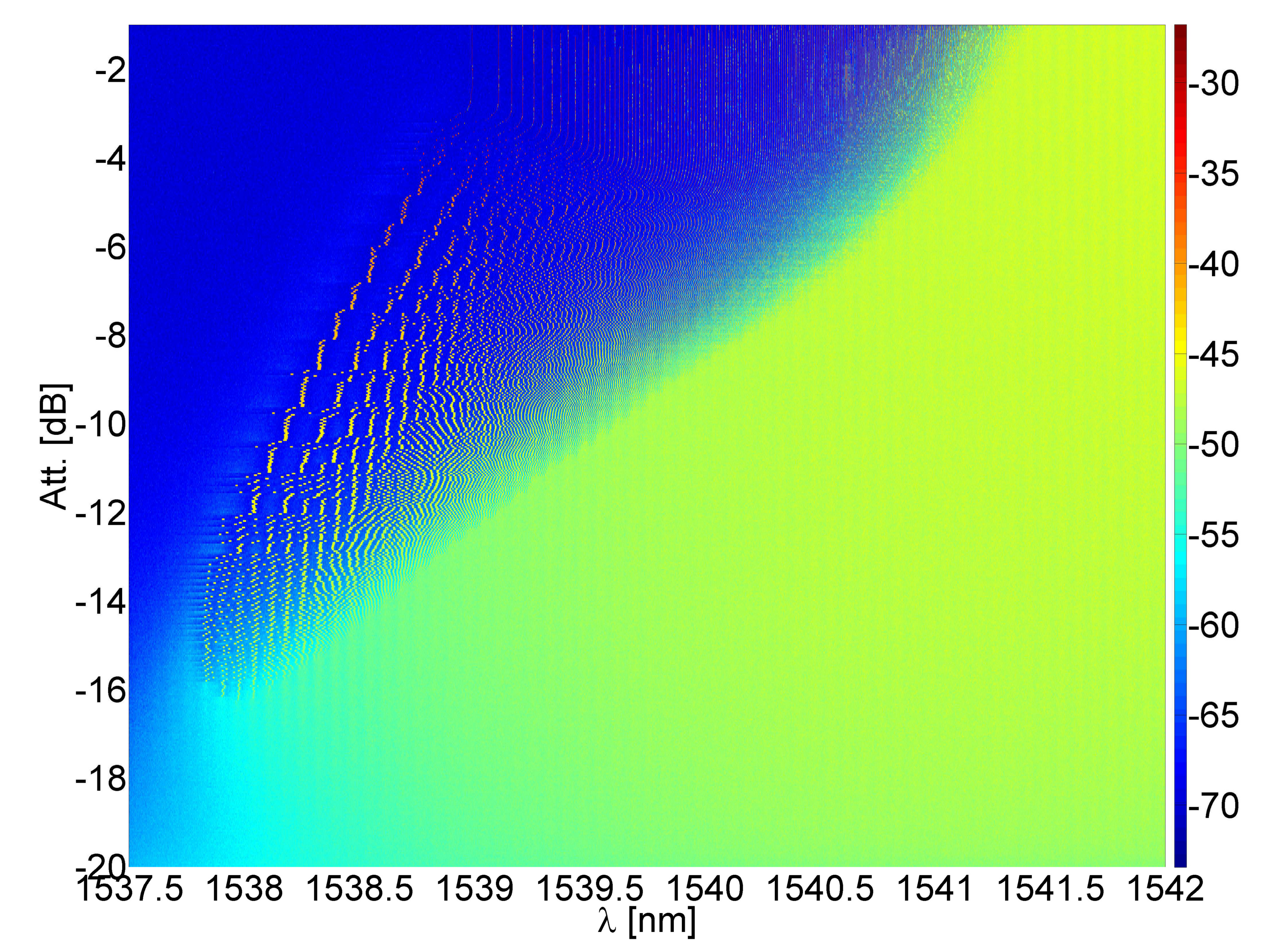}
\end{center}
\caption{In loop attenuator. The EDF length is $l_{\mathrm{EDF}}=5
\operatorname{m}$, the temperature is $T=2.8 \operatorname{T}$, and diode
current is $I_{\mathrm{D}}=150\operatorname{mA}$. Attenuation is changed from
high ($-20$ dB) to low ($-1$ dB).}%
\label{FigAtt}%
\end{figure}

Manipulating the effective and normalized EDF length $\eta_{\mathrm{EDF}}$ by
changing the loop loss coefficient $\alpha_{\mathrm{FL}}$ is explored below in
the next section (see Fig. \ref{FigAtt}), whereas in the current section
$\eta_{\mathrm{EDF}}$ is controlled by exploiting the temperature dependence
of the emission length $l_{\mathrm{E}}$ (see Fig. \ref{FigSetupT}). As the
temperature is lowered the emission length $l_{\mathrm{E}}$ increases, and
consequently the effective and normalized EDF length $\eta_{\mathrm{EDF}}$
decreases. In the short EDF limit, for which $\eta_{\mathrm{EDF}}\ll1$, it is
expected from the condition $\mathrm{d}\varrho/\mathrm{d}\lambda=0$\ [see Eq.
(\ref{varrho})] that lasing will occur in the band where emission cross
section peaks near $\lambda_{\mathrm{g}}\simeq1540%
\operatorname{nm}%
$ \cite{Desurvire_547,Desurvire_246,Zyskind_869,Hickernell_19}. This behavior
is demonstrated by the plot shown in Fig. \ref{FigSetupT}(b) for temperatures
below $10%
\operatorname{K}%
$. The overlaid red rectangle in Fig. \ref{FigSetupT}(b) indicates the region
shown in Fig. \ref{FigSetupT}(c) with higher resolution, inside which USOC occurs.

\textbf{Attenuator} - To explore the dependency on the loop loss
$\alpha_{\mathrm{FL}}$ coefficient, a tunable attenuator is integrated into
the loop [in the location of the X-box shown in Fig. \ref{FigSetupT}(a)]. The
plot in Fig \ref{FigAtt}, which displays the measured optical spectrum as a
function of attenuation, demonstrates the strong USOC dependence on loop loss
$\alpha_{\mathrm{FL}}$. This dependency diminishes as the wavelength increases.

The condition $\mathrm{d}\varrho/\mathrm{d}\lambda=0$ implies in the short EDF
limit that $\left(  \mathrm{d}\lambda_{\mathrm{g}}/\mathrm{d}\alpha
_{\mathrm{FL}}\right)  ^{-1}=-\alpha_{\mathrm{FL}}\log\alpha_{\mathrm{FL}%
}\left(  \log l_{\mathrm{f}}^{\prime}\right)  ^{\prime}$, where prime denotes
a derivative with respect to wavelength $\lambda$ [see Eq. (\ref{varrho})].
This relation together with the experimental results shown in Fig.
\ref{FigAtt} yield the value of $1/\left(  \log l_{\mathrm{f}}^{\prime}\right)
^{\prime}=0.3%
\operatorname{nm}%
$. Note that this value is about $10$ times smaller than the value of
$1/\left(  \log l_{\mathrm{f}}^{\prime}\right)  ^{\prime}$ that is extracted
from room temperature absorption and emission measurements that are reported
in Ref. \cite{Franco_1090} for a similar EDF.

\begin{figure}[ptb]
\begin{center}
\includegraphics[width=3.2in,keepaspectratio]{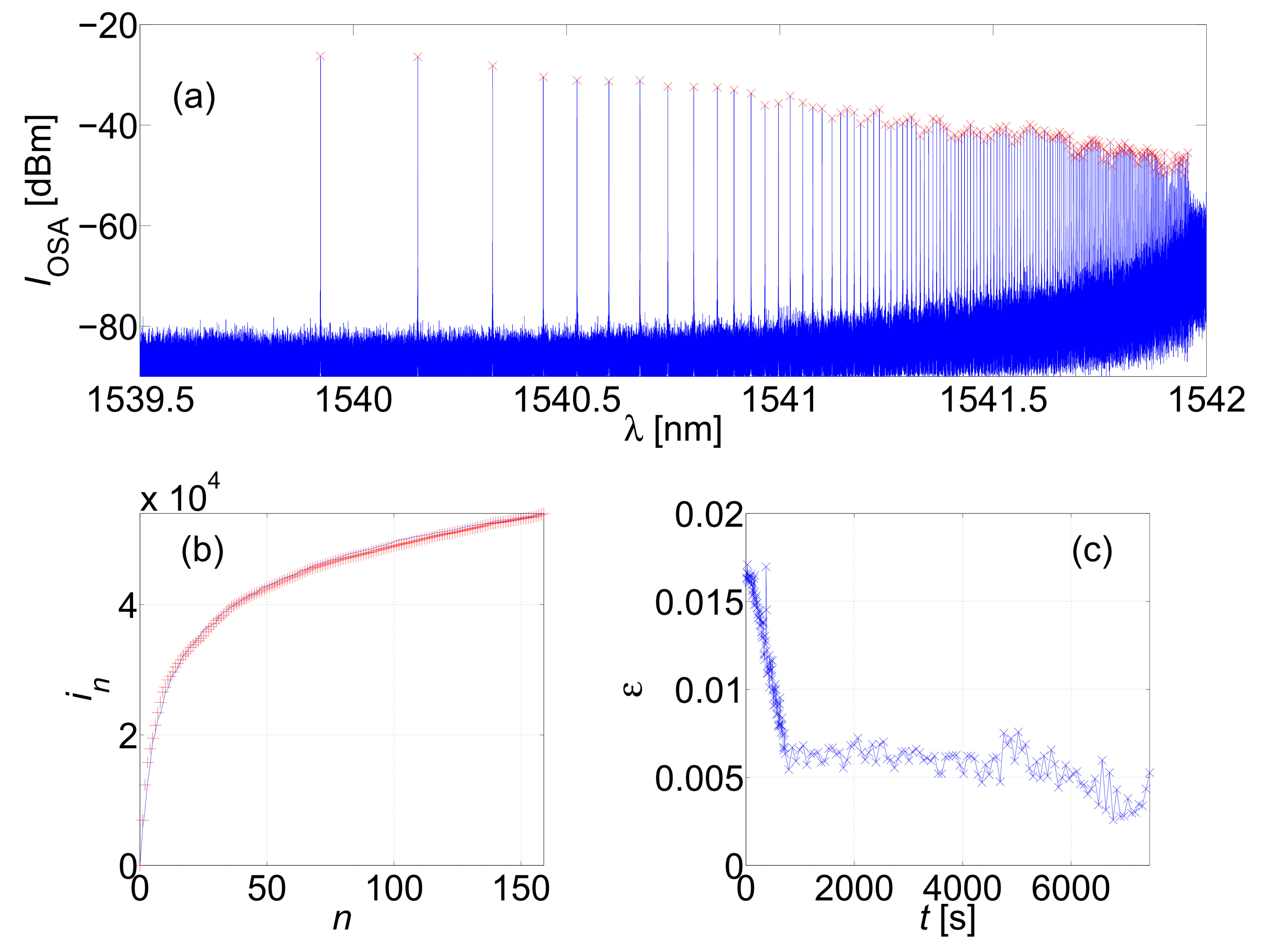}
\end{center}
\caption{{}USOC. (a) OSA signal $I_{\mathrm{OSA}}$ as a function of wavelength
$\lambda$. EDF length is $l_{\mathrm{EDF}}=10 \operatorname{m} $, diode
current is $I_{\mathrm{D}}=200 \operatorname{mA} $, and temperature is
$T=3.2\operatorname{K}$. (b) The measured values of normalized frequency detuning $i_{n}=\left(  f_{0}
-f_{n}\right)  /f_{\mathrm{L}}$ (red cross symbols), and the calculated values
of $\nu\log p_{n}$ with $\nu=7888$ (blue solid line) [see Eq. (\ref{i_n})].
(c) The level of disagreement $\varepsilon$
[of fitting using Eq. (\ref{i_n})] as a function of
the elapsed time $t$ after switching on the diode current.}%
\label{FigUSOC}%
\end{figure}

\textbf{USOC} - The USOC $n$'th peak wavelength is denoted by $\lambda_{n}$,
where $n=0,1,2,\cdots$ [see Fig. \ref{FigSetupT}(c) and Fig. \ref{FigUSOC}%
(a)]. The frequency $f_{n}$ associated with the $n$'th peak is given by
$f_{n}=c/\lambda_{n}$, where $c$ is the speed of light in vacuum, and the
corresponding normalized frequency detuning $i_{n}$ is defined by $i_{n}%
\equiv\left(  f_{0}-f_{n}\right)  /f_{\mathrm{L}}$, where $f_{\mathrm{L}%
}=c/\left(  n_{\mathrm{F}}l_{\mathrm{L}}\right)  $ is the loop frequency,
$n_{\mathrm{F}}=1.45$ is the fiber refractive index, and $l_{\mathrm{L}}$ is
the fiber loop total length.

The detuning sequence $\left\{  i_{n}\right\}  $ is found to be well describe
by the following empirical law \cite{Buks_128591}%
\begin{equation}
i_{n}=\nu\log p_{n}\ , \label{i_n}%
\end{equation}
where $\nu$ is a positive constant, and $p_{n}$ is the $n$'th prime number.
The comparison between the measured values of $i_{n}$ and the calculated
values of $\nu\log p_{n}$ [see Eq. (\ref{i_n})] yields a good agreement [see
Fig. \ref{FigUSOC} (b)]. The level of disagreement is quantified by the
parameter $\varepsilon=n_{\mathrm{m}}^{-1}\sum_{n=1}^{n_{\mathrm{m}}%
}\left\vert \left(  i_{n}-\nu\log p_{n}\right)  /i_{n_{\mathrm{m}}}\right\vert
$, where $n_{\mathrm{m}}$ is the number of peaks that can be reliably
resolved. The level of disagreement $\varepsilon$ is shown in Fig.
\ref{FigUSOC} (c) as a function of the elapsed time $t$ after switching on the
diode current. The plot reveals that during the first 10 minutes after
switching on the disagreement parameter $\varepsilon$ rapidly drops down
towards a value of about $0.005$.

As can be seen from the empirical law given by Eq. (\ref{i_n}),
all frequency spacings $f_{n^{\prime}}-f_{n^{\prime\prime}}$ between pairs of
USOC peaks are unique [recall the fundamental theorem of arithmetic, and that
$\log x+\log y=\log\left(  xy\right)  $]. As is discussed below, this
observation suggests a connection between USOC formation and intermode
coupling. The loop optical complex amplitude at time $t$ is expressed as
$g\left(  f_{\mathrm{L}}t\right)  $, where the function $g\left(  x\right)  $
is Fourier expanded as $g\left(  x\right)  =\sum_{n=-\infty}^{\infty}%
c_{n}e^{2\pi inx}$, where $c_{n}$ is the complex amplitude of the $n$'th loop
mode. When dispersion can be disregarded, the time evolution of the (assumed
slowly varying) amplitude $c_{n}$ is governed by $\dot{c}_{n}=-\partial
_{n}^{\ast}\mathcal{H}+\xi_{n}$, where $\partial_{n}$ denotes the Wirtinger
derivative for the $n$'th mode, and where the terms $\xi_{n}$ represent noise
\cite{Haus_1173}. The contribution of intermode coupling to the total
Hamiltonian function $\mathcal{H}$ is given by $\mathcal{H}_{\mathrm{c}%
}=\left(  \gamma^{\prime\prime}v_{\mathrm{g}}/2\right)  \left\Vert
g\right\Vert _{4}^{4}$, where $\left\Vert g\right\Vert _{p}$ denotes the
$L_{p}$ norm of $g$. Note that $\left\Vert g\right\Vert _{2}^{2}$ is the total
optical intensity, and that $\left\Vert g\right\Vert _{4}^{4}=\sum_{n^{\prime
}-n^{\prime\prime}+n^{\prime\prime\prime}-n^{\prime\prime\prime\prime}%
=0}c_{n^{\prime}}^{{}}c_{n^{\prime\prime}}^{\ast}c_{n^{\prime\prime\prime}%
}^{{}}c_{n^{\prime\prime\prime\prime}}^{\ast}$. It was shown in Ref.
\cite{Buks_128591} that for a fixed $\left\Vert g\right\Vert _{2}^{2}$ (total
optical intensity), $\left\Vert g\right\Vert _{4}^{4}$ is nearly minimized
when USOC is formed, provided that $\gamma^{\prime\prime}>0$ (i.e. RSA occurs
in the USOC band). This observation suggests a connection between USOC
formation and RSA, however, it has remained unclear how the assumption that
dispersion can be disregarded can be justified.

Typically, USOC is highly stable and nearly temperature independent below about 
$5\operatorname{K}$ [see Fig. \ref{FigSetupT}(b)]. In contrast, strong temperature dependency and significant drift in time of USOC peaks are observed at higher temperatures.

Further insight can be gained from the degree of polarization
(DOP) of light emitted from the output port. DOP is measured by integrating a
rotating quarter wave plate with the coherent optical spectrum analyzer. Using
this method we find a DOP value of about $0.1$ in both USOC and in the
continuous (above $1542%
\operatorname{nm}%
$) lasing bands. Note that commonly DOP is below about $0.15$ in fiber lasers,
unless a polarization-maintaining fiber is used \cite{Liem_CMS4}.

\begin{figure}[ptb]
\begin{center}
\includegraphics[width=3.2in,keepaspectratio]{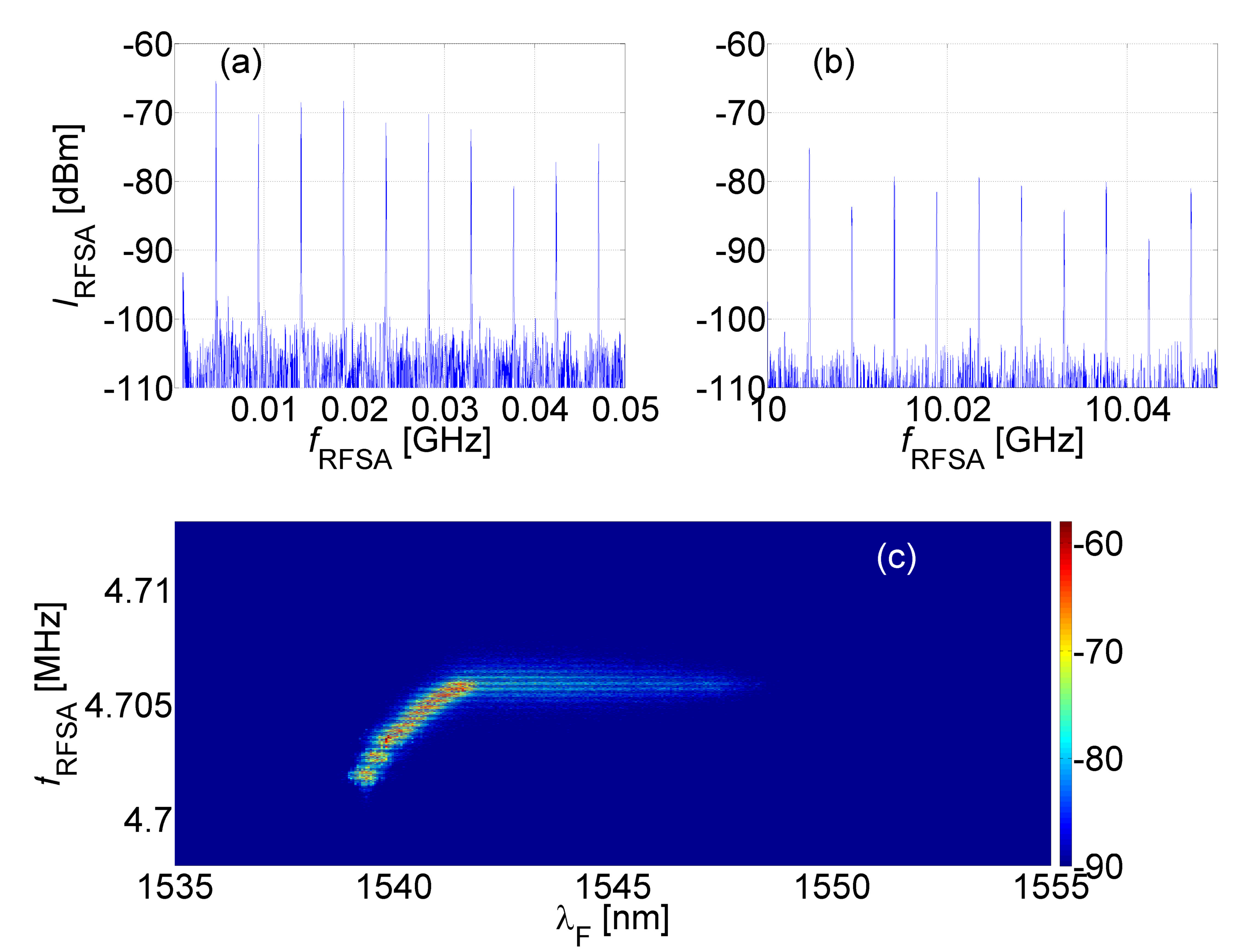}
\end{center}
\caption{{}RFSA. The measurements presented in (a) and (b) are performed
without the tunable OF. EDF length is $l_{\mathrm{EDF}}=10 \operatorname{m} $,
diode current is $I_{\mathrm{D}}=200 \operatorname{mA} $, and temperature is
$T=4.1 \operatorname{K} $. (c) The measured first harmonic peak (in dBm units)
as a function of the tunable OF wavelength $\lambda_{\mathrm{F}}$.}%
\label{FigRFSA}%
\end{figure}

\textbf{RF spectrum} - Further insight can be gained from probing the PD
signal using a RFSA [see Fig. \ref{FigSetupT}(a)]. The RF spectrum shown in
Fig. \ref{FigRFSA}(a) and (b), which is taken \textit{without} the tunable OF
[see Fig. \ref{FigSetupT}(a)], contains an equally spaced comb of peaks
centered at frequencies $n_{\mathrm{L}}f_{\mathrm{L}}$, where $f_{\mathrm{L}%
}=4.705%
\operatorname{MHz}%
$ is the loop frequency, and where $n_{\mathrm{L}}$ is a positive integer
(peaks with $n_{\mathrm{L}}\lesssim2500$ can be resolved). The measured first
harmonic peak (i.e. $n_{\mathrm{L}}=1$) is shown in Fig. \ref{FigRFSA}(c) as a
function of the tunable OF wavelength $\lambda_{\mathrm{F}}$. This plot
reveals that the\ measured value of the loop frequency $f_{\mathrm{L}}$ inside
the USOC band ($\lambda_{\mathrm{F}}\lesssim1542%
\operatorname{nm}%
$) drops well below its measured value in the continuous band ($\lambda
_{\mathrm{F}}\gtrsim1542%
\operatorname{nm}%
$).

The observed drop of $f_{\mathrm{L}}$ in the USOC band can be generally
attributed to both chromatic and nonlinear dispersion. The contribution of
chromatic dispersion can be estimated from the wavelength dependency of the
EDF absorption coefficient $\alpha_{\mathrm{EDF}}$ \cite{Le_3611}. The
Kramers-Kr\"{o}nig formula, which relates $\alpha_{\mathrm{EDF}}$ and the EDF
refractive index $n_{\mathrm{EDF}}$\ through a Hilbert transform
\cite{Hutchings_1,Sai_136,Hickernell_19,Montagna_871,Arkwright_798}, can be
used to calculate $n_{\mathrm{EDF}}$, which, in turn, yields the corresponding
group index \cite{Gehring_3752}. Comparing the calculated group index to the
data shown in Fig. \ref{FigRFSA}(c) indicates that the observed drop in
$f_{\mathrm{L}}$ is far larger and far sharper compared to what is
theoretically expected (when only chromatic dispersion is taken into account).
This observation leads to the conclusion that the contribution of chromatic
dispersion to the observed drop is relatively small. In other words, the
dominant underlying mechanism responsible for the observed drop in
$f_{\mathrm{L}}$ inside the USOC band is nonlinear dispersion, which is
characterized by the real part $\gamma^{\prime}$ of the nonlinear coefficient
$\gamma$.

\textbf{EDF gain} - Open loop measurements are performed in order to
characterize the EDF gain $g_{\mathrm{EDF}}$, and intermodulation response
(see next section). In these measurements the loop is opened in the location
of the X-box shown in Fig. \ref{FigSetupT}(a).

\begin{figure}[ptb]
\begin{center}
\includegraphics[width=3.2in,keepaspectratio]{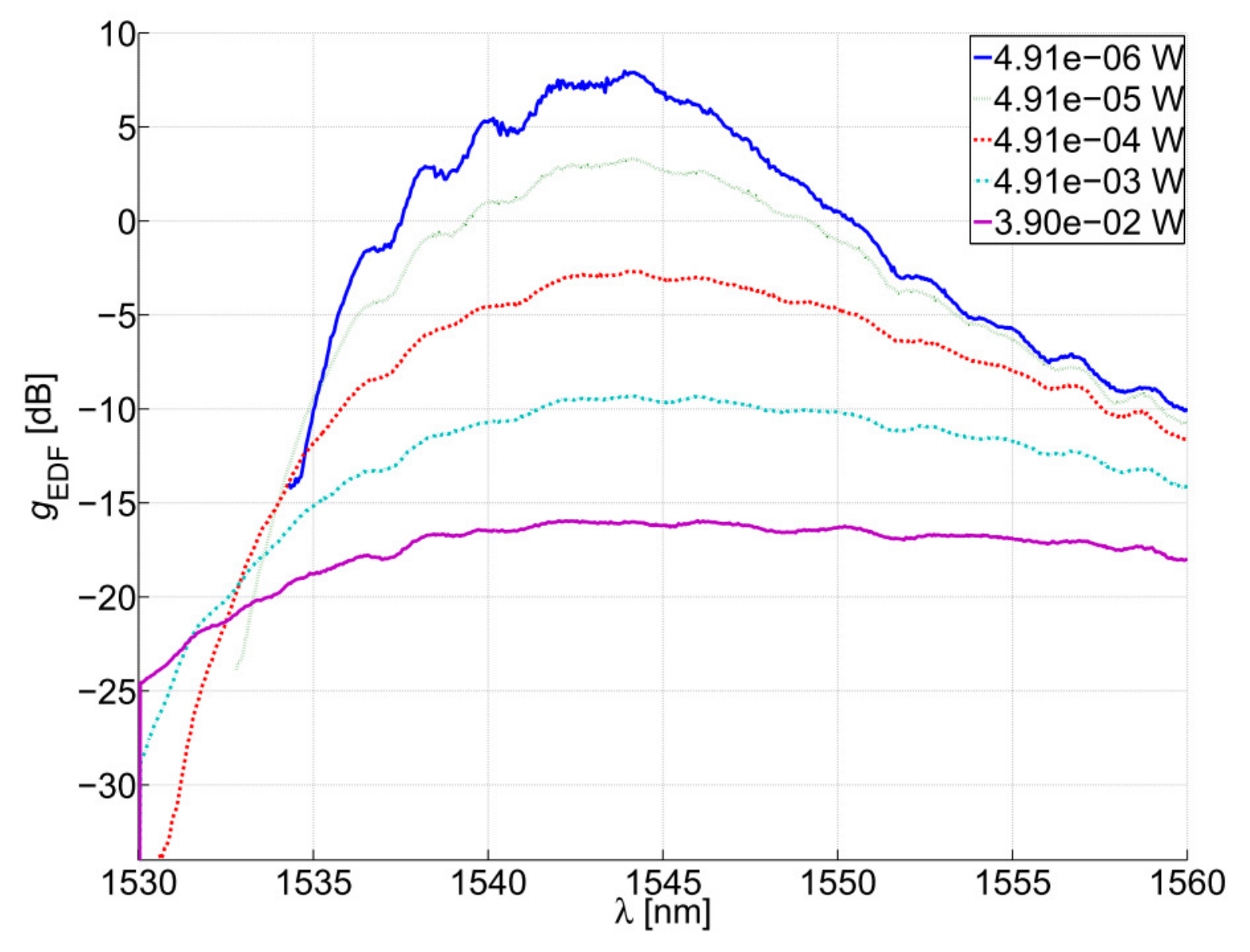}
\end{center}
\caption{{}EDF gain. EDF length is $l_{\mathrm{EDF}}%
=10\operatorname{m}$, diode current is $I_{\mathrm{D}}=150\operatorname{mA}$,
and temperature is $T=2.9\operatorname{K}$. Open loop EDF gain
$g_{\mathrm{EDF}}$ as a function of wavelength $\lambda$ for different values
of the laser input power $P_{\mathrm{in}}$ (values of $P_{\mathrm{in}}$ are
indicated by the figure legend).}%
\label{FigOpenLoopGain_vs_P}%
\end{figure}

To measure the EDF gain, a narrow band laser having a tunable
wavelength $\lambda_{\mathrm{L}}$ is employed. Open loop EDF gain
$g_{\mathrm{EDF}}$ is plotted in Fig. \ref{FigOpenLoopGain_vs_P} as a function
of wavelength $\lambda$ for different values of the laser input power
$P_{\mathrm{in}}$. For the band $\lambda\gtrsim1535\operatorname{nm}$, the
measured EDF gain $g_{\mathrm{EDF}}$ decreases as the optical input power
$P_{\mathrm{in}}$ is increased (see Fig. \ref{FigOpenLoopGain_vs_P}). This
observation suggests that RSA and gain saturation occur in this wavelength
band (inside which USOC is observed when the loop is closed).

\begin{figure}[ptb]
\begin{center}
\includegraphics[width=3.2in,keepaspectratio]{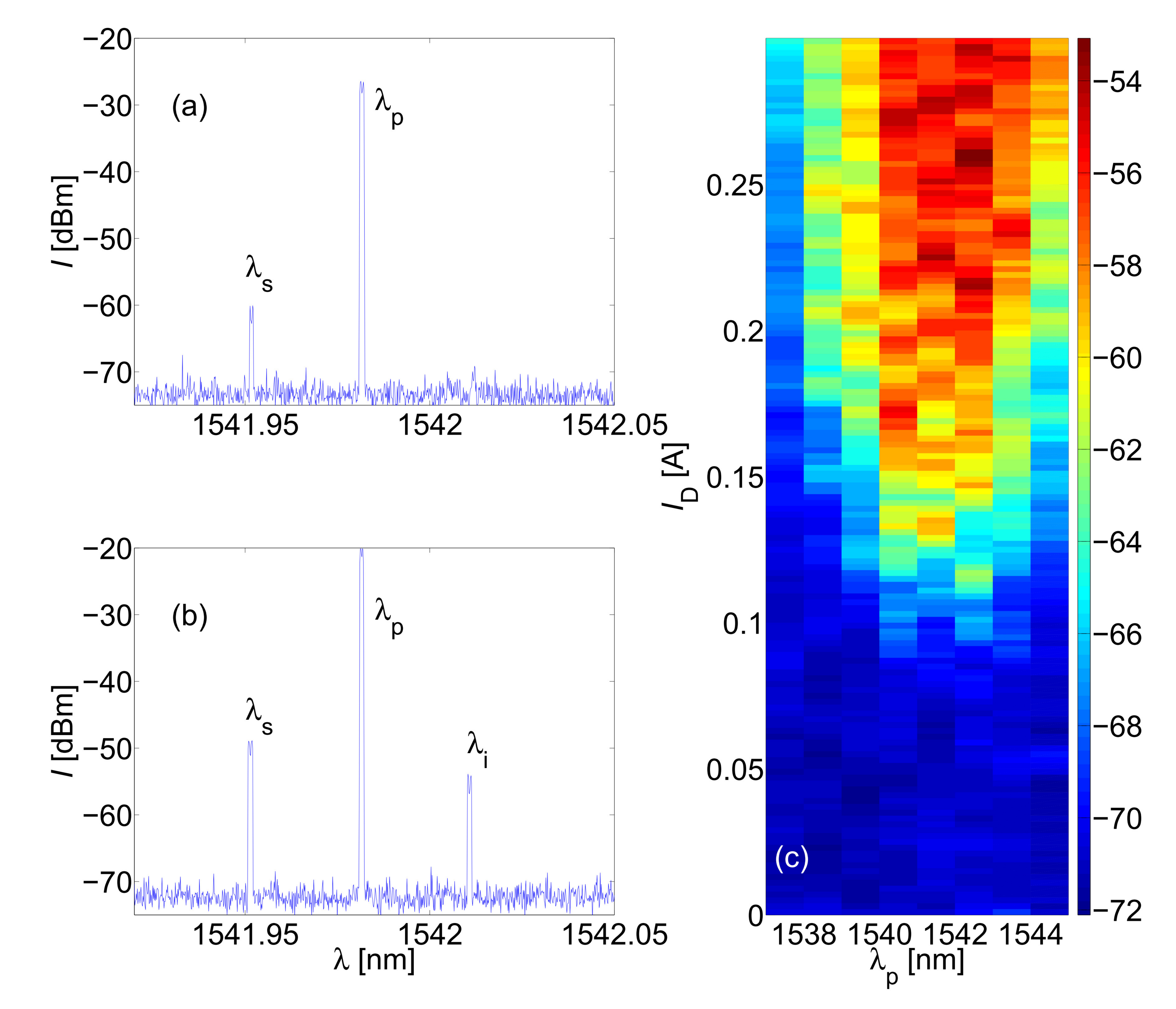}
\end{center}
\caption{Intermodulation gain - EDF length is $l_{\mathrm{EDF}}%
=10\operatorname{m}$, and temperature is $T=2.9\operatorname{K}$. (a) The
two-tone input spectrum generated using a driven ferrimagnetic sphere
resonator \cite{Nayak_193905}. (b) The output spectrum. (c) The idler
(wavelength $\lambda_{\mathrm{i}}$) peak intensity
(in dBm units) as a function of pump wavelength $\lambda_{\mathrm{p}}$ and
diode current $I_{\mathrm{D}}$.}%
\label{FigIMD}%
\end{figure}

\textbf{Intermodulation} - The technique of intermodulation is employ to
characterize intermode coupling. In this technique two (phase locked)
monochromatic tones are injected into the system under study (i.e. the EDF).
The first one, which has a relatively large amplitude, and a wavelength
denoted by $\lambda_{\mathrm{p}}$, is commonly refereed to as the
\textit{pump}. The second one is a relatively low-amplitude \textit{signal}
tone having wavelength $\lambda_{\mathrm{s}}=\lambda_{\mathrm{p}}%
-\lambda_{\mathrm{d}}$, where $\lambda_{\mathrm{d}}$ is the detuning
wavelength, which is assumed to be small $\left\vert \lambda_{\mathrm{d}%
}\right\vert \ll\lambda_{\mathrm{p}}$. A driven ferrimagnetic sphere resonator
is employed for the generation of this two-tone input signal
\cite{Nayak_193905} [see Fig. \ref{FigIMD}(a)].

Frequency mixing between the pump and signal input tones occurring in the EDF
gives rise to an \textit{idler} tone at the output [see Fig. \ref{FigIMD}(b)].
The intensity of the idler peak at wavelength $\lambda_{\mathrm{i}}%
=\lambda_{\mathrm{p}}+\lambda_{\mathrm{d}}$ is shown in Fig. \ref{FigIMD}(c)
as a function of pump wavelength $\lambda_{\mathrm{p}}$ and diode current
$I_{\mathrm{D}}$. The results presented in Fig. \ref{FigIMD}(c) suggest that
intermode coupling is relatively strong in the region where USOC occurs.

The intermodulation gain $g_{\mathrm{I}}\equiv P_{\mathrm{i}%
}/P_{\mathrm{s}}$, is the ratio between the generated idler power
$P_{\mathrm{i}}$ and the input signal power $P_{\mathrm{s}}$ \cite{Yurke_5054}%
. The intermodulation gain $g_{\mathrm{I}}$ is expected to depend on the
nonlinear dispersion coefficient $\gamma^{\prime}$. Measurements of EDFs
similar to the ones used in the current study yield values for $\gamma
^{\prime}$ in the range of $2-30%
\operatorname{W}%
^{-1}%
\operatorname{km}%
^{-1}$ (at room temperature, and in the telecom band)
\cite{Agrawal_Nonlinear_fiber_optics,shaulov1998modeling,Roy_6791}. On the
other hand, when nonlinear absorption $\gamma^{\prime\prime}$ is disregarded,
the extraction of $\gamma^{\prime}$ from the measured intermodulation gain
$g_{\mathrm{I}}$ [see Fig. \ref{FigIMD}(c)] and the EDF gain $g_{\mathrm{EDF}%
}$ (see Fig. \ref{FigOpenLoopGain_vs_P}) yields values for $\gamma^{\prime}$
at least one order of magnitude larger (in the band where USOC occurs). This
observation suggests that other nonlinear processes (e.g. nonlinear
absorption) play an important role in the optical band where USOC is formed
(when the loop is closed).

\textbf{In loop OF} - Nonlinear dispersion tends to broaden lasing linewidth
\cite{Roy_6791,Lapointe_232,Babin_1729,Feng_155}. To explore this, an OF
having a tunable wavelength $\lambda_{\mathrm{F}}$, and linewidth of
$\delta_{\mathrm{F}}=1.2%
\operatorname{nm}%
$, is integrated into the fiber loop [in the location of the X-box shown in
Fig. \ref{FigSetupT}(a)].

\begin{figure}[b]
\begin{center}
\includegraphics[width=3.2in,keepaspectratio]{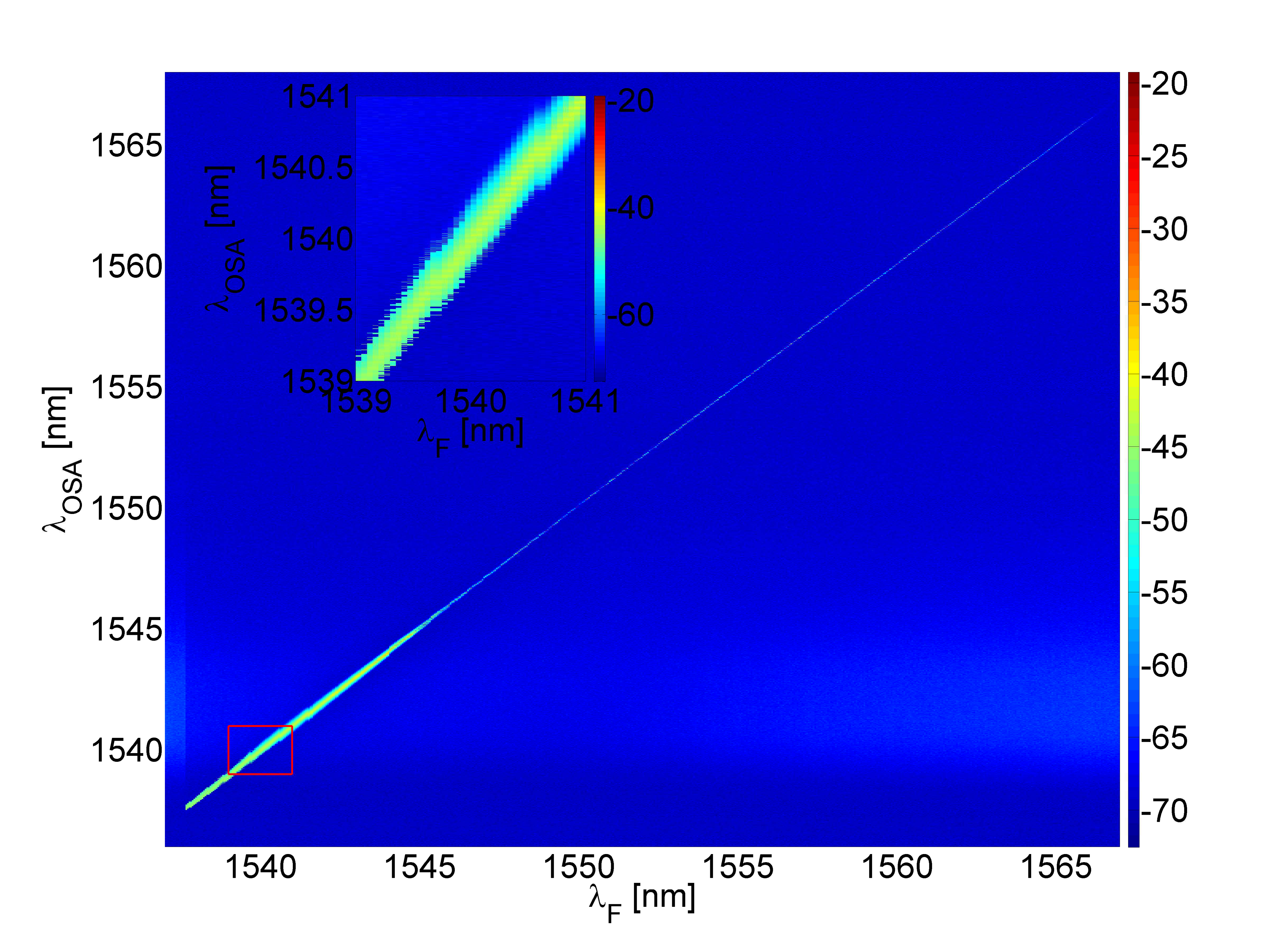}
\end{center}
\caption{{}In loop OF. EDF length is $l_{\mathrm{EDF}}=10\operatorname{m}$,
diode current is $I_{\mathrm{D}}=120\operatorname{mA}$, and temperature is
$T=2.9\operatorname{K}$.}%
\label{FigILOF}%
\end{figure}

The lasing linewidth $\delta_{\mathrm{L}}$ is related to the
nonlinear dispersion coefficient $\gamma^{\prime}$ by
\begin{equation}
\delta_{\mathrm{L}}=\sqrt{\gamma^{\prime}l_{\mathrm{L}}P_{\mathrm{C}}}%
\delta_{\mathrm{F}}\;,\label{delta_L}%
\end{equation}
where $P_{\mathrm{C}}$ is the intra-cavity optical power \cite{Lapointe_232}. This
relation allows the extraction of $\gamma^{\prime}$ from the measured lasing
linewidth $\delta_{\mathrm{L}}$. The measured optical spectrum is shown in
Fig. \ref{FigILOF} as a function of $\lambda_{\mathrm{F}}$. The region where
$\lambda_{\mathrm{F}}$ is tuned to the band where USOC is formed (in the
absence of the OF) is shown in higher resolution in the inset of Fig.
\ref{FigILOF}. The measured lasing linewidth $\delta_{\mathrm{L}}$ is about $2%
\operatorname{pm}%
$ in the range $\lambda_{\mathrm{F}}\in\left[  1548%
\operatorname{nm}%
,1565%
\operatorname{nm}%
\right]  $, whereas $\delta_{\mathrm{L}}\simeq200-500%
\operatorname{pm}%
$ for $\lambda_{\mathrm{F}}$ in the band where USOC occurs (in the absence of
the OF) [see Fig. \ref{FigSetupT}(b)]. The relation (\ref{delta_L}) yields the
value of $\gamma^{\prime}=1.4%
\operatorname{W}%
^{-1}%
\operatorname{km}%
^{-1}$ for the range $\lambda_{\mathrm{F}}>1548\operatorname{nm}$, whereas
significantly larger values for $\gamma^{\prime}$ are obtained for the band
where USOC occurs. This observation further supports the hypothesis that other
nonlinear processes play an important role in the band where USOC is formed [note that nonlinear absorption is disregarded in the derivation of Eq. (\ref{delta_L})].

\begin{figure*}[ptb]
\begin{center}
\includegraphics[width=6.4in,keepaspectratio]{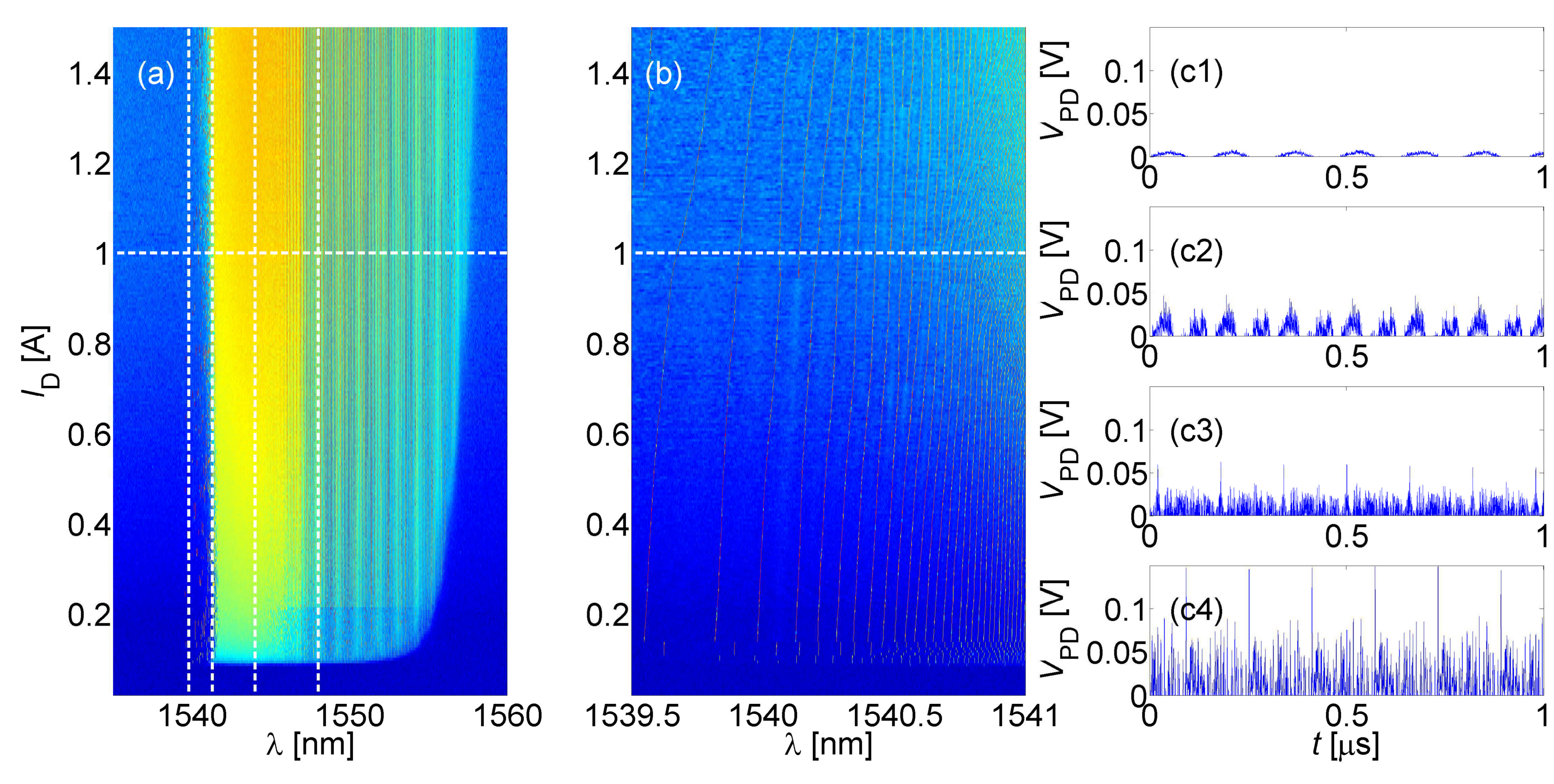}
\end{center}
\caption{Graphite SA. EDF length is $l_{\mathrm{EDF}}=5 \operatorname{m}$, and
temperature is $T=3.2 \operatorname{T}$. OF wavelength [labeled by vertical
dotted lines in (a)] is (c1) $\lambda_{\mathrm{F}}=1539.8 \operatorname{nm}$,
(c2) $\lambda_{\mathrm{F}}=1541.3 \operatorname{nm}$, (c3) $\lambda
_{\mathrm{F}}=1544 \operatorname{nm}$ and (c4) $\lambda_{\mathrm{F} }=1548
\operatorname{nm}$, and diode current [labeled by horizontal dotted lines in
(a) and (b)] for the time traces shown in (c1), (c2), (c3) and (c4) is
$I_{\mathrm{D}}=1000 \operatorname{mA}$.}%
\label{FigSA_G}%
\end{figure*}

\textbf{SA} - To further explore the role played by nonlinear absorption
(characterized by the imaginary part $\gamma^{\prime\prime}$ of the nonlinear
coefficient $\gamma$), a SA made of graphite \cite{Lin_880,Lin_045109} is
integrated into the fiber loop [in the location of the X-box shown in Fig.
\ref{FigSetupT}(a)]. The measured optical spectrum is shown as a function of
diode current $I_{\mathrm{D}}$ in Fig \ref{FigSA_G}(a). The USOC region is
shown in higher resolution in Fig \ref{FigSA_G}(b).

The effect of the integrated SA is explored by measuring PD time traces for
different values of the OF wavelength $\lambda_{\mathrm{F}}$ [see Fig.
\ref{FigSA_G}(c)]. Note that the OF is installed outside the loop [see Fig. \ref{FigSetupT}(a)]. The periodic short pulses that are shown in Fig.
\ref{FigSA_G}(c3) and (c4) indicate that ML occurs in the continuous band
($\lambda_{\mathrm{F}}\gtrsim1542%
\operatorname{nm}%
$). On the other hand, inside the USOC band ($\lambda_{\mathrm{F}}\lesssim1542%
\operatorname{nm}%
$) no pulses are observed [see Fig. \ref{FigSA_G}(c1) and (c2)], thus the USOC
band does not participate in the ML. The observed coexistence of USOC and ML
suggests that the coupling between the USOC and continuous bands is relatively
weak. This weakness is partially attributed to the above-discussed
drop in the loop frequency $f_{\mathrm{L}}$ that is observed in the USOC band
[see Fig. \ref{FigRFSA}(c)], which, in turn, gives rise to phase mismatch
\cite{Shibata_1205,schneider2004nonlinear}, and suppression of coherent
coupling between the USOC and continuous bands
\cite{Taheri_013828,Godey_063814,Mcmillan_411}.

\textbf{Discussion} - Further study is needed to reveal the
underlying mechanisms responsible for USOC formation. On the one hand, a
simple theoretical model that disregards dispersion, suggests a connection
between USOC formation and RSA. On the other hand, our experimental findings
invalidate the assumption that dispersion can be disregarded. Theoretical
modeling of the system under study is highly
challenging in the presence of dispersion, because explicit time dependency cannot be eliminated by a
transformation into a rotating frame. Future work will be devoted for the
development of a theory capable of quantitatively reproducing the highly
rich and convoluted behaviors that are experimentally observed.

\textbf{Acknowledgments} - Useful discussions with Luca Leuzzi, Vassilios
Kovanis and Kerry Vahala are acknowledged. We thank Michael Shlafman for his help in graphite preparation.

\textbf{Disclosures} - The authors declare no conflicts of interest.

\bibliographystyle{ieeepes}
\bibliography{acompat,Eyal_Bib}

\end{document}